# Eliminating Scanning Delay using Advanced Neighbor Discovery with Caching (ANDWC)

Hassan Keshavarz (Corresponding author)

Department Of Computer Systems & Technology,

University Of Malaya, 50603 Kuala Lumpur, Malaysia

Tel: (+60)-172-950-923   E-mail: keshavarz_hassan@IEEE.org

Ghaith A. Abdulwahed

Department Of Computer Systems & Technology,

University Of Malaya, 50603 Kuala Lumpur, Malaysia

E-mail: ghaithalrawi2@gmail.com

Rosli Salleh

Department Of Computer Systems & Technology,

University Of Malaya, 50603 Kuala Lumpur, Malaysia

E-mail: rosli_salleh@um.edu.my

Laeth A. Abdul Wahid

Department Of Computer Systems & Technology,

University Of Malaya, 50603 Kuala Lumpur, Malaysia

E-mail: laithalrawi80@yahoo.com

**Abstract**

The advance in wireless technologies and portable devices such as smart phones has made the Wireless Local Area Networks (WLANs) popular in the recent years. Nowadays, WLANs have become widely accepted in both private and public sectors due to ease of installation, reasonable prices and high data rates that can support real time applications. However, fast handoff required for such real-time applications is not provided in the current IEEE 802.11 specifications. Consequently, providing seamless mobility in these WLANs is an important issue. To solve this problem, a new fast handoff scheme called Advanced Neighbor Discovery with Caching (ANDWC) is proposed. This new mechanism is based on the user's mobility between two or more different Basic Service Sets (Layer 2 mobility). ANDWC can eliminate scanning delay (which contributes up to 90% of the total handoff latency) to provide seamless handoff by using pre-neighbor-discovery and caching mechanisms.

**1. Introduction**

WLANs have become very popular in recent years as they are easy to build and configure, inexpensive compared to wired networks and they allow users to move around within an acceptable coverage area and still be connected to the network with a high speed link of up to 54 Mbps (802.11 based WLANs) [3]. Consequently, real time and multimedia applications such as voice and video over Internet Protocol (VOIP) have emerged and been supported by this type of wireless LANs.  However, as the access points (APs) that build the WLANs have a relatively short coverage range (50 – 300 m) [17], many APs are needed to expand the WLAN. In this case, a mobile user has to change association with an old AP to a new one (perform a handoff) if moving and the signal strength from the user's associated AP gets weaker in order to keep the connection to the network and access its available resources [20]. Such a process (handoff) may, unfortunately, affect the previously mentioned real-time applications if it takes more than 50 ms (which is the case in IEEE 802.11), as no data will be sent or received from the network. As a result, many ongoing researches focus on improving handoff delay so that uninterruptable VoIP can be supported. The remainder of this paper is outlined as follows. Section 2 introduces the IEEE802.11. Section 3 elaborates IEEE802.11 Hand-off procedure, Section 4 describes Hand-off Delay and Section 5 briefly reviews related work. The proposed scheme is presented in Section 6 , while Section 7 discuss the result   of the proposed scheme, respectively, followed by the conclusion presented in Section 8.

## 2. IEEE 802.11

*2.1IEEE 802.11 standards*

There are currently three well known wireless physical layer standards: 802.11a, 802.11b and 802.11g [1, 2]. The 802.11a standard works in the 5 GHz Industrial Scientific and Medical (ISM) band with 32 channels offered, 8 of which do not overlap. This standard uses Orthogonal Frequency Division Multiplexing (OFDM) modulation and offers a data rate of up to 54 Mbps. The 802.11b/g standards operate in the 2.4 GHz ISM band. They use 11 channels offered by the 2.4 GHz spectrum (in the USA). Only 3 of these channels (1, 6 and 11) do not overlap. Similar to 802.11a, the 802.11g standard uses OFDM. ODFM allows the 802.11g to offer data rates of 54 Mbps. On the other hand, 802.11b utilizes the Direct Sequence Spread Spectrum (DSSS) modulation method which offers maximum data rates of up to 11 Mbps. Since the 2.4 GHz has longer wave length than the 5 GHz band, its signals offer both 802.11b/g and naturally longer ranges than 802.11a. IEEE 802.11f, or the Inter Access Point Protocol (IAPP), is another IEEE 802.11 extension. The IAPP facilitates the communication and secure transfer of all context information relevant to a Mobile Station (MS) between multi vender APs during a handoff [10]. The context information of a Mobile Station (MS) carried by the Move-Response packet may consist of a timestamp, BSSID, sequence number, etc [5, 10, 12].

*2.2 IEEE 802.11 mobility types*

Current research in wireless mobility [7, 11] can be divided into three types: No-transition, Basic Service Set (BSS) -transition (micro mobility or Layer 2 mobility) and Extended Service Set (ESS)-transition (macro mobility or Layer 3 mobility) depending on the layers involved in this operation (Figure 1). In No-transition, local STA movement within the coverage area of the AP can occur. BSS -transition (micro mobility) is the capability to move between APs within the same subnet or routed domain (within one ESS). This type of mobility (Layer 2) needs cooperation between the APs in order to exchange information necessary for a successful handoff in which the IAPP can be used, and is the type that this project focuses on. Finally, ESS-transition (macro mobility) is the ability to move between APs located in different (two or more) ESSs belonging to different IP subnets which may be administrated by different organizations like two companies, etc. This type of mobility introduces IP addressing issues (Layer 3).

## 3. IEEE802.11 Handoff Procedure

The complete handoff procedure can be divided into two distinct logical phases: scanning and re-authentication phases.

*3.1 Scanning phase*

Scanning is a MAC layer mechanism performed by an MS to determine the surrounding APs to which it can re-associate [17]. It can be divided into three sub-phases: detection, scanning and decision.

*3.2 Detection sub-phase*

During this sub-phase, an MS decides whether or not to start scanning, based on different factors such as failed frame transmissions or the detection of a weak signal from the associated AP. In other words, the Received Signal Strength (RSS) is below a certain threshold [2, 8, 15, 23].

3.2.1 Scanning sub-phase

After scanning is detected, an MS starts scanning all the available channels (11 in 802.11b/g and 32 in 802.11a) trying to reach any APs in range and associate to one of them [18]. Some studies claim that the scanning phase is responsible for over 90% of the total handoff latency [17], because the MSs must wait for a specific time on each channel (depending on the scan type used). Consequently, the overall scanning delay may be costly (350-500 ms for active scan and 1100-3200 ms for passive scan) [13]. IEEE 802.11 defines two scanning modes: passive and active scanning [2], something that allows manufacturers to choose between low power consumption and fast handoff.

In passive scanning, an STA must listen to beacon frames periodically broadcasted by APs. This can be done by switching to each channel in order to know the parameters of those APs, such as beacon interval, capability information, BSSID, supported rate, etc. [14,22]. Although this type of scanning requires minimal power consumption, it is time consuming at about1100-3200 ms [2, 20]. In active scanning, an MS actively broadcasts probe request messages to all APs on all available channels and waits for response messages from those APs. By doing so, the MS can obtain the same system information it can get from hearing a periodic beacon in passive scanning [17]. The duration each channel is probed for is based on two variables, namely, the MinChannelTime and MaxChannelTime, the minimum and maximum channel times. When an MS actively scans a channel, it broadcasts a probe request frame and begins the probe timer. The MS waits for probe responses. If 'no' is received by MinChannelTime, the MS assumes the channel is empty and it switches to scan the next channel. If one or more responses are heard within the MinChannelTime, the MS must wait for the duration of the MaxChannelTime and processes all of the responses received by this time. This is to allow the APs in range to reply and be considered as next candidates.

3.2.2 Decision sub-phase

During the decision sub-phase, an MS determines which AP it should connect to [4]. The RSS and available bandwidth are some parameters the decision depends on. If the difference between the RSSI of the new AP and RSSI of the old AP exceeds a pre-specified value [12], the STA will handoff to the new AP.

*3.3 Re-authentication phase*

The two sub–phases normally involved in the re-authentication phase are Re-authentication and Re-association [14]. In Re-authentication sub-phase, the new AP either accepts or rejects the MS identity. This process starts by sending a re-authentication request frame from the MS to the new AP passing on the MS identity. The AP then sends back a re-authentication response frame to the MS as well as the response of acceptance or rejection. The two sub-types of authentication defined by IEEE 802.11 standard are [16].Open system authentication - no authentication is required and the AP always accepts the MS. This type of authentication uses only the authentication request and authentication response frames.Shared Key authentication based on wired equivalent privacy (WEP) is when a four-way authentication mechanism is used (four messages are exchanged) and in The re-association sub-phase starts once the new AP has successfully authenticated the MS [16]. This process involves the exchange of a re-association request and response frame between the MS and the new AP, after which the association of the MS will be moved from the old AP to the new AP. The handoff process is now completed.

**4. Handoff Delay**

The entire handoff delay is the combination of Scanning and the Re-authentication delays.

*4.1 Scanning delay*

The scanning delay can vary depending on which scan mode is used (passive or active/fast mode) [16]. If passive scanning is used, the average delay can be represented as a function of the beacon interval and the number of channels available which introduces long delay, since an MS may not know the arrival time of beacons. Consequently, the MS has to stay on the channel for a beacon interval. For instance, if the beacon interval is 100 ms, the average scanning delays of IEEE 802.11b/g with 11 channels is 100x11=1100 ms and 802.11a with 32 channels is 100x32=3200 ms, with no respect to the channel switching time since it can be very small.If, however, active scanning is

used,average scanning delay can be determined by the number of channels to scan, and the time to stay on a channel and wait for the response messages (MinChannelTime and MaxChannelTime values) [17]. Although MinChannelTime and MaxChannelTime values are device-specific [18], it is reported by Yong [2006] that MinChannelTime is about 20 ms, and MaxChannelTime is about 40ms. So the scanning delay (SD) can be: $N \times MinChannelTime \leqslant SD \leqslant N \times MaxChannelTime$. Where N is the number of channels to be scanned. According to Jeng-ji [2007], active scanning delay may take between 350 and 500ms. Since scanning delay in the active and passive scanning depends on the number of channels; using fast scan, or selective scanning, can reduce delay because the number of channels probed/listened to is less, since only a selected set of channels are probed/listened to [16]. For example, only non overlapping channels (1, 6 and 11) will be probed using active fast scan rather than all 11 channels in 802.11b.

*4.2 Re-authentication delay*

The re-authentication delay includes both re-authentication and re-association delays [18].

4.2.1 Re-authentication delay

This latency is incurred during the authentication frames exchange. The re-authentication process consists of two or four consecutive frames depending on the authentication method used by the AP (open system or shred key authentication). Experimental results Sangheon [2007] show that the re-authentication sub-phase using open system authentication may take about 1-2ms.

4.2.2 Re-association delay

This type of latency is incurred during the re-association frames exchange between the MS and the AP. Experimental results Sangheon [2007] show that the re-association sub-phase may take about 1-2ms. However, if the IAPP is used in the network, additional IAPP messages between the old and new AP will be involved in order to transfer MS credentials from the old to the new AP. Consequently, the re-association latency can increase up to 40ms.

**5. Related Work**

Yazan [2008] proposed a new advanced handoff mechanism for 802.11 networks called First Satisfaction First Reservation Handoff (FSFR-HO). Unlike the typical handoff mechanism, this new method requires the MS to scan the channels until it finds the first AP with an RSS value above a predefined Satisfactory Threshold (ST). Then the MS immediately enters the re-authentication phase without continuing to scan the remaining channels. The ST in this method is set equal to the handoff threshold value (Satisfactory Threshold (ST) = Handoff Threshold (HT)). Although this new mechanism reduces scanning delay by reducing the number of channels to be scanned, it may lead to inappropriate handoff decisions as the MS will choose the first next AP found on the scanned channel depending on the handoff threshold and not on the best RSS, which can result in weak connection or even connection loss, especially if the MS is moving away from the newly selected AP.Another method for reducing handoff delay in WLANs is proposed Hye-Soo [2006], where the authors use an improved AP with an additional radio frequency (RF) module called a SNIFFER. This SNIFFER is used for monitoring the movement of the MS by enabling the improved AP to listen (eavesdrop) to the medium access control (MAC) frames of the incoming MS, and then obtaining the address of the old AP associated with that MS. After that, the information needed for the re-association to the new AP is transferred to the STA through the old AP. In this proposed method, modifying the AP by adding a new interface introduces extra hardware costs, especially if there is a large number of APs in the network.

**6. Our proposed Mechanism**

In this section, we first identify some of the network assumptions for our new mechanism, then we identify the newly created IAPP packets and some new management frames along with their duties. Following that, the mechanism is explained by dividing it into multiple phases and figures are shown. Finally, simulation results from OMNeT++ will be presented for both the new mechanism and the IEEE802.11b/g standard to show the contribution of our proposed mechanism.

*6.1 Network assumptions*

  i. The following are some general assumptions for our network.
  ii. Mutual receiving coverage between APs.
  iii. High speed LAN is built between APs (Full Duplex).
  iv. Each AP knows about all other APs on LAN (New IAPP Packets are presented for this purpose).
  v. Each AP knows which APs are its neighbours (IEEE802.11 APInfo frames are used for this purpose).
  vi. MS will re-associate to the best neighbour in the HANDOFF list.
  vii. MS moves linearly along a straight path (Linear Mobility).

*6.2 New involved packets and frames*

We have created new IAPP packets and management frames needed during the implementation phase of our new Mechanism.

6.2.1 New IAPP packets

The following four new IAPP packets are added as extensions to the original IAPP and will be used for communication between the APs on LAN during and after the INFORMATION COLLECTION phase. They will have command numbers starting from 7 (normal IAPP packets occupy command numbers from 0-6 that identify their specific function).APInfoREQ Packet: Used by each AP to send an AP information request to all APs living on LAN (Destination MAC address will be the IAPP multicast address 224.0.1.178). The IAPP command number is 7.

APInfoRES Packet: Used by each AP to send its information back (e.g. SSID, MAC address, IP address, Beacon Interval, channel number, etc.) to the caller AP (AP that sent APInfoREQ). Its IAPP command number is 8.PERMISSION Packet: Used to send permission to the nextMAC AP in order to start its AIR VERIFICATION phase. The IAPP command number is 9. NEW_NEIGBOUR_REQ Packet : Used when a new AP is discovered by an MS during the STATION ASSOCIATION PHASE. It will be sent to the old AP to update its APInfo List. Its IAPP command number is 10. NEW_NEIGBOUR_RES Packet: Used when NEW_NEIGBOUR_REQ Packet is received. It will be sent to the newly discovered AP in order to update its APInfo List. The IAPP command number is 11.

6.2.2 IEEE 802.11APinfo frame

This new management frame will be used during the AIR VERIFICATION phase and it has four sub type frames as follows:

IEEE 802.11APInfoREQ Frame: Used in the AIR VERIFICATION phase to send a request to a neighbouring AP available in the APInfo List.

IEEE 802.11APInfoRES Frame : Used in the AIR VERIFICATION phase to send neighbour information (e.g. Received signal strength, etc.) back to the caller AP.

IEEE 802.11APInfoACK Frame: Used to acknowledge the receipt of information from the neighbour and also to send the information of the AP verifying its neighbour to the neighbouring AP to modify its APInfo List.

IEEE 802.11 APInfoUpdate Frame: Used by MS to send the information of its old AP to the new AP discovered during a full scan.

Modified Beacon Frame: The original beacon frame will be modified to hold the neighbour list of an AP along with the AP information. This frame is normally sent to the MS.

Original Authentication and Association Frames will be used to connect the MS to the new AP during the STATION ASSOCIATION PHASE. Advanced Neighbour Discovery with Caching (ANDWC)

Our mechanism (ANDWC) is divided into three phases:

6.2.3 Information collection phase (done on LAN)

This phase is divided into two sub phases, namely, COLLECTION and VOTING. Both have two probable scenarios: ALL ON in which all the APs on LAN are performing the information and voting sub-phases at the same time, and NEWADDED in which new APs are added after these two sub-phases have finished.

6.2.3.1 All on scenario

During the COLLECTION sub phase, information about all APs on LAN is exchanged and saved in each AP (Figures 2 & 3). This sub-phase begins as soon as the AP is turned on, when each AP also starts by sending IAPP APInfoREQ packets to all other APs within the multicast domain asking them to send their information (AP uses IAPP multicast address 224.0.1.178). The AP then waits for responses. Upon receiving the IAPP APInfoREQ packet, each AP will send its information back to the caller AP using IAPP APInfoRES packet. This information may include any of the following: SSID, BSSID, MAC address, IP address, channel number, or beacon intervals, etc. Whenever the caller AP receives these responses, it collects this information and pushes it into an AP Information List (APInfo List).

After collecting all information on all the APs on LAN, the collector AP assumes that all APs in the APInfo List are neighbours. Only after that, the VOTING sub-phase starts. In the VOTING sub-phase, each AP builds a MAC LIST in which all MAC addresses of all AP Wireless Network Interface cards (WNICs), including the MAC address of the AP that builds the MAC LIST, will be saved and sorted in ascending order. This MAC LIST will be used for the voting process in which all APs will vote for the AP with the lowest MAC address to start the AIR VERIFICATION phase. Each AP identifies its position in the MAC LIST. By doing this, an AP can also discover which AP will start the AIR VERIFICATION after itself by setting the nextMAC value equal to the next bigger MAC address in the MAC LIST. Only after the voting sub-phase is complete, the AP with the lowest MAC address starts the AIR VERIFICATION phase while the other APs wait for that AP to finish and grant them permission to start their AIR VERIFICATION.

6.2.3.2 Air verification phase (done on WLAN)

This phase also has two probable scenarios, the ALL ON in which all the APs on LAN finish voting together and start the AIR VERIFICATION PHASE, and NEWADDED in which a new AP is added after the voting sub-phase is finished.

6.2.3.2.1 All on scenario

In this phase, the AP with the lowest WNIC's MAC address enters the AIR VERIFICATION PHASE by sending an IEEE 802.11APInfoREQ frame to all APs on the APInfo list, one by one (Figures 2 & 3). The caller AP switches to each channel in the APInfo list and sends IEEE 802.11 APInfoREQ using MAC information obtained during the INFORMATION COLLECTION phase. If no response is received, the AP will re-transmit that frame. After one re-transmission, this neighboring AP will be marked as a non real neighbor by setting the real Neighbor = false and the caller AP switches to verify the next neighbor in the list. Upon frame reception by the neighboring AP, it measures the signal strength (RSS) of the received frame and send this information back to the caller AP inside an IEEE 802.11APInfoRES frame. The caller AP then marks this AP as a real neighbor in its APInfo List by setting the real Neighbor = true, adds its received signal strength to its record and copies the information of this neighbor into a REAL_NEIGHBOUR list. The caller AP also measures the RSS of the IEEE 802.11APInfoRES frame from the neighboring AP. Next, the caller AP acknowledges receipt of information by sending back an IEEE802.11APInfoACK frame to the neighboring AP which includes the RSS (RSS of Response frame). Upon receiving the ACK frame, the neighboring AP also marks the caller AP as a real neighbor and pushes it into its REAL_NEIGHBOUR list too. By using IEEE802.11APInfo sub frames, we can acknowledge received information at the caller side and reduce overall AIR VERIFICATION phase time by reducing the messages exchanged between neighboring APs.

This is done by using a mutual verification process to verify the correctness of the two APs' information simultaneously instead of repeating the process at each AP. This verification process is done on all APs in the APInfo list not marked as real neighbors, until the end of the list. Finally, after all APs in the list are verified and only real neighbors are marked (set realNeighbor=true), the AP checks the I_AM_THE_LAST value which is a variable indicating that the AP is the last in the AIR VERIFICATION sequence. If I_AM_THE_LAST=true, the AP realizes it is last in the AIR VERIFICATION sequence and no more APs are waiting for its PERMISSION, so it sets nextMAC=NULL and starts sending beacons.

If I_AM_THE_LAST=false (default when all APs are turned on), the AP will check the nextMAC value and get the IP address of the same AP which has this MAC address and send an IAPP PERMISSION packet to this IP address, and then it starts sending beacons which include only real neighbors. When an AP receives the IAPP

PERMISSION packet, it checks the value of the I_AM_THE_LAST included in the packet; if I_AM_THE_LAST=false, then starts the same AIR VERIFICATION phase and sends PERMISSION to nextMAC when finished, else set nextMAC=NULL and start the AIR VERIFICATION Phase.

We have decided to use the PERMISSION packet along with an AIR VERIFICATION because if the AIR VERIFICATION phase is not done sequentially, in an orderly manner, the situation becomes very messy and the neighbor verification process may lead to incorrect information. Let's look at the scenario bellow. If there were three APs (AP1 with channel no.1, AP2 with channel no.6 and AP3 with channel no.11) on the LAN and each wanted to start the AIR VERIFICATION phase without any pre-allocated order, then, if AP1 wanted to start first, it would check the information stored in the APInfo list and start by switching to the first channel in the list (Ch. 6) and send an IEEE 802.11 APInfoREQ frame. If AP2 with Channel no. 6 also wanted to start verification it would switch to the first channel in the APInfo list (Ch. 1) and also start sending an IEEE802.11 APInfoREQ frame. This, neither AP discovers the other but instead they mark each other as unreal neighbors (set realNeighbour=false). Therefore, to solve this problem we use an automatic voting process based on MAC addresses of WNICs (similar to the voting process in switches for choosing the Root Bridge (RB) in the STP).

6.2.3.2.2 New added AP scenario

In this scenario, we explain the details of adding a new AP onto the LAN. As all the APs might have finished their information collection phase but not the AIR VERIFICATION phases, this scenario has added more difficulty to the design process of our mechanism. It is probable that a newly added AP has finished the INFORMATION COLLECTION and the VOTING sub-phases and it has discovered that the MAC address of its WNIC is less than the

LAST AP, or it has the lowest MAC in the MAC LIST. Thus it will find that it has the permission to start the AIR VERIFICATION process while at the same time there may be another AP (old AP) at the same phase. In such a scenario the new AP may lead to the problem discussed in the previous sub-section (incorrect neighbour verification). In order to solve this problem, we have decided to use the newAP and the I_AM_THE_LAST fields in both the IAPP APInfoRES packet and the IAPP PERMISSION packet (on LAN). These two fields will force any newly added APs to wait for a PERMISSION to arrive from the last AP (with a nextMAC value equal to NULL) in the previously built MAC list (during the last voting process) in order to start its AIR VERIFICATION phase.

The whole process is performed in the following way. When a new AP is added into the LAN, it checks the APInfo list. If the APinfoList is empty, then this AP is new and it starts the normal INFORMATION COLLECTION phase by sending an IAPP APInfoREQ packet and then waits for responses to arrive from all APs on LAN (multicast domain). At this time, the newly added AP assumes it will start the voting process with all other APs. However, when each AP receives the IAPP APInfoREQ packet, it checks the I_AM_THE_LAST value. If I_AM_THE_LAST equals to false, then the receiver AP will send back its information in an IAPP APInfoRES packet to the caller AP (normal process for non last APs). If I_AM_THE_LAST equals true, then this AP discovers that it is the last in the MACList and that the caller AP is newly added. This old AP must then inform the caller AP that it is new. So the last AP (which has the nextMAC=NULL or I_AM_THE_LAST=true) will modify its information by setting the nexMAC equal to the MAC address of caller AP and I_AM_THE_LAST equal to false.

It will also check the FINISHED value (FINISHED=false when all APs are turned on). The FINISHED value indicates that the AP has already finished its AIR VERIFICATION phase. If the receiver of the IAPP APInfoREQ packet finds that FINISHED equals false, then it will set the newAP field in the IAPP APInfoRES packet = true (new AP must wait) and send the response to the caller AP with no PERMISSION. If the receiver AP finds that FINISHED=true, then it sets the newAP=true and sends an IAPP APInfoRES packet followed by an IAPP PERMISSION packet to the nextMAC AP (using the IP address of that AP saved in the APInfo list). Whenever the caller

AP receives the IAPP APInfoRES packets from all APs on LAN it will check the newAP field in each packet. If newAP equals to false, then it will just add the received AP information and wait for more responses. When the last response packet arrives, the new AP checks the newAP field and discovers that it is new. Consequently, it will set I_AM_THE_LAST value = true (I am newly added) and wait for PERMISSION to arrive from the old AP. When an IAPP PERMISSION packet arrives, the newly added AP will set the nextMAC=NULL and finally it will start its AIR VERIFICATION phase. By doing so, we have guarantee that no new AP (even with a lower MAC address) can destroy the previously defined sequence during the VOTING sub-phase (Figure 3).

6.2.3.3 Station association phase (WLAN)

Finally, when AIR VERIFICATION is completed and all APs have discovered their real neighbours along with the neighbours' RSSs, the APs sort their REAL_NEIGHBOUR lists depending on the strongest RSS value (strongest RSS first). For instance, If AP2 has two neighbours, namely AP1 and AP3, with RSSI = 3 and 4 respectively, then its REAL_NEIGHBOUR list will have AP3, AP1 even if AP1 was verified before AP3 during the AIR VERIFICATION phase. By doing so, each AP reduces the load on the MS by sorting its REAL_NEIGHBOUR list instead of sorting it by each MS. Another benefit of sorting the REAL_NEIGHBOUR list is that the strongest RSSs mean that neighbouring APs are nearest to this AP, so the AP can guarantee that MS will choose the nearest/strongest RSSI AP for it. After sorting the list the AP starts sending it to each beacon (Figure 4).

Whenever a beacon is received by an MS, it saves the REAL_NEIGHBOUR list in a Handoff list used by the MS to associate itself to the best real neighbouring AP in range. When the RSSI from the associated AP drops below a pre-specified threshold, the MS triggers handoff by sending a RE-AUTHENTICATION REQUEST to the first AP in the Handoff list. If the MS receives a RE-AUTHENTICATION RESPONSE from that AP, it measures the RSS from the new AP and compares it with the SAVED RSS (RSS for the same AP saved in the Handoff List). If the NEW RSS $\geq$ SAVED RSS, the MS directly associates to the new AP by sending it an ASSOCIATION REQUEST. Else, if NEW RSS < SAVED RSS, or if there is no acknowledgment from the AP after one corresponding retransmission, the MS switches to the next AP in the Handoff List and tries to authenticate itself with that AP. The MS has to switch to the next AP in the Handoff list because no acknowledgment from the previous AP in the list indicates that this neighbour AP is not in the MS coverage range (moving direction).

If the MS reaches the end of the Handoff list, or the REAL_NEIGHBOUR list in the beacon was received with no appropriate AP to associate itself with, the MS switches to full scan mode by scanning all available channels hoping that a good received signal (RSS) from a new AP can be detected. The MS does a full scan only if it enters a different ESS with APs not yet informed by the beacons from the associated AP. If a new AP is detected during the full scan process, or an AP marked as a non real AP in the APInfo list of the old AP responds, the MS will associate itself with this new AP and then send an IEEE 802.11APInfoUpdate frame to the new associated AP. Upon receiving this frame the new AP sends an IAPP NEW_NEIGHBOUR_REQ packet to the old AP to mark it as a real neighbour into its APInfo list (the new AP will find the IP address of the old AP in the IEEE 802.11APInfoUpdate frame which is received from MS). When the old AP receives an IAPP NEW_NEIGHBOUR_REQ packet, it adds the caller AP as a real neighbour into its APInfo and REAL_NEIGHBOUR lists and sends an IAPP NEW_NEIGHBOUR_RES packet back to the caller AP.

By following this new caching mechanism, each AP guarantees that every MS in coverage range will choose the best neighbour to associate itself with no matter what the MS's position is in the coverage circle. It also guarantees that any new neighbouring AP discovered during the STATION ASSOCIATION phase will update both the new and old APs APInfo and REAL_NEIGHBOUR lists which can be considered an extension to the pre-neighbour discovery mechanism that can support layer three handoffs.

## 7. SIMULATION RESULTS

Here, we present simulation results for the total AIR_VERIFICATION phase time and handoff time using full scan (standard) and ANDWC along with packet loss measurement collected using OMNeT++ simulation.

### 7.1 Air verification phase time

When the AP starts it collects information on all APs on LAN and verifies this information on the WLAN. As long as the information collection and voting phases are done on high speed LAN, the average latency introduced is not big. However, switching to each channel in the APInfo list and verifying each neighbour takes longer. Figure 5 shows that by using a mutual verification process the overall AIR_VERIFICATION phase can be reduced by verifying two APs simultaneously. This continuously reduces air verification time as the last APs' turn comes on the list.

### 7.2 Handoff time

The results obtained are presented in Figure 6. Clearly, using ANDWC can eliminate 90% of the handoff latency introduced by the full scan phase when an MS directly enters the authentication phase by sending an IEEE802.11APInfoREQ frame to the first MAC address in the HANDOFF list . Many trials performed show the average handoff latency is 2 ms if the MS authenticates itself with the first AP in the HANDOFF list (without wrong neighbour information). However, if an MS tries to authenticate itself with the first AP in the HANDOFF list and there is no response, the MS switches to the next MAC address in the HANDOFF list, introducing extra delay of 1 ms average added to the normal handoff latency (2 ms). Clearly, handoff latency using ANDWC depends on the number of wrong neighbours supplied by the associated AP (in beacons of old AP). Figure 6 shows if the first MAC is not in the MS's motion direction, the delay is 3 ms (average) and if the first two APs are not in the MS's moving direction, then latency is 4 ms (average), a huge improvement from total latency measured using full scan (standard) of 304 ms .

*7.3 Packet loss*

To measure packet loss, UDP packets were transmitted to an MS to simulate a video stream during Handoff. Using full scan and ANDWC mechanisms, we could measure the packet loss of 12500 packets sent to the MS from a video server. Clearly, using ANDWC reduces packet loss from 188 (using full scan) to only 6 packets resulting from the lower latency of ANDWC during the handoff.

## 8. CONCLUSION

This paper covers the fundamentals of IEEE 802.11 based WLANs by reviewing some well-known standards used. It also describes the mobility types that may occur and lead to triggering Handoff. As this paper focuses on the handoff process occurring due to micro mobility (Layer 2), it covers the three phases of handoff process and the delays imposed by each phase. Some recent mechanisms for reducing the scanning delay during a handoff are reviewed and advantages and disadvantages identified. Finally, a new, fast mechanism for reducing handoff delay to a level that can support multimedia applications is presented with some new IAPP packets and management frames.

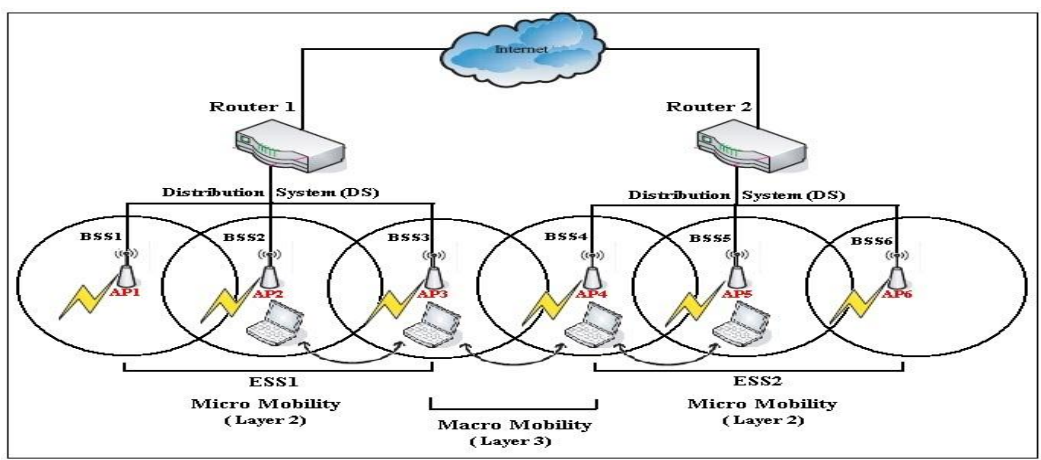

Figure 1. IEEE 802.11 Wireless Mobility

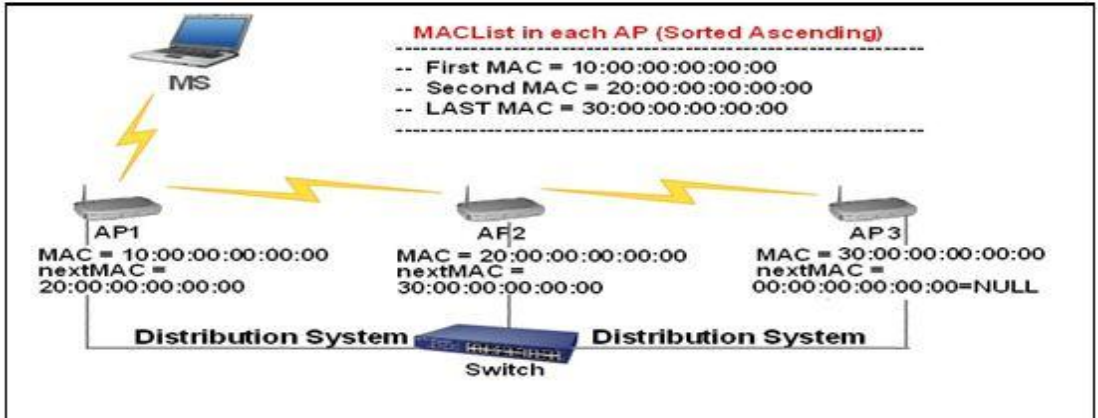

Figure 2. Information Collecting & Air Verification Phases

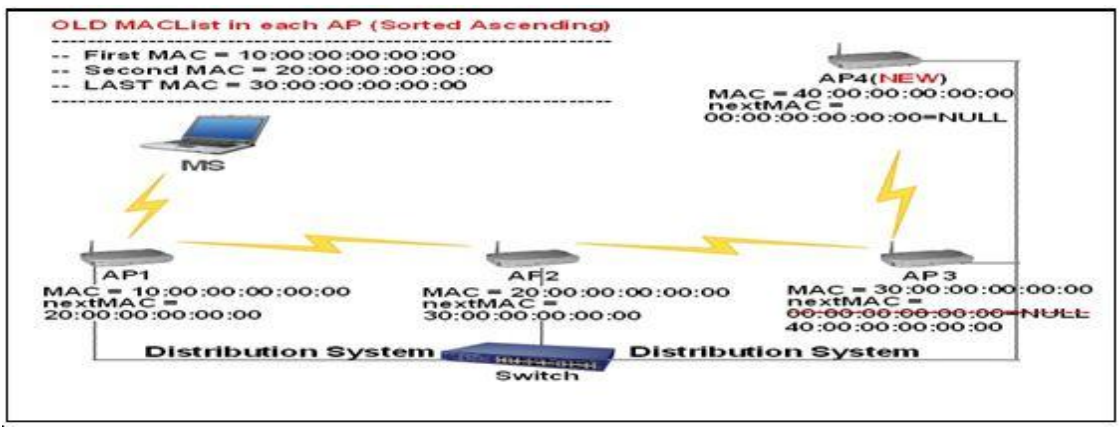

Figure 3. New Added AP

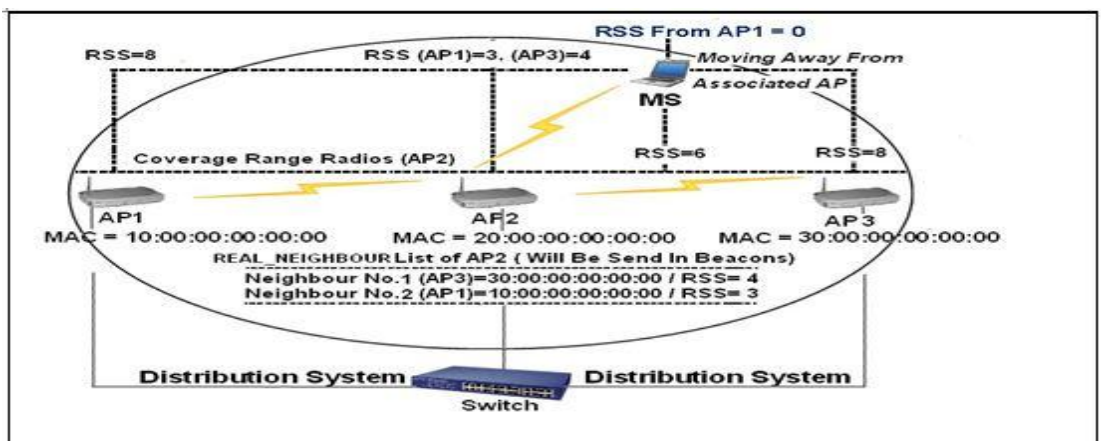

Figure 4. Station Association Phase

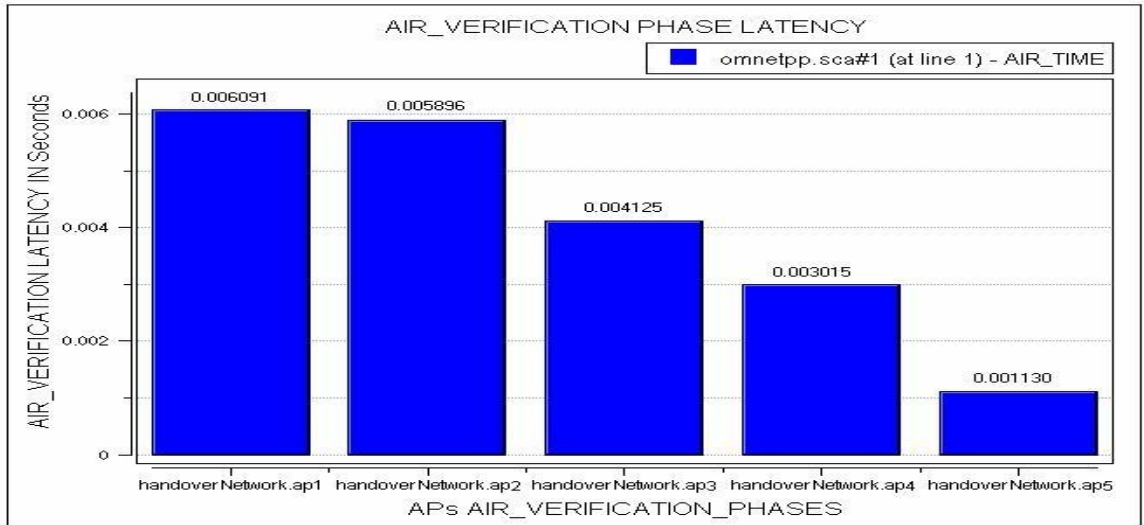

Figure 5. Air_Verificatio Phase Latency

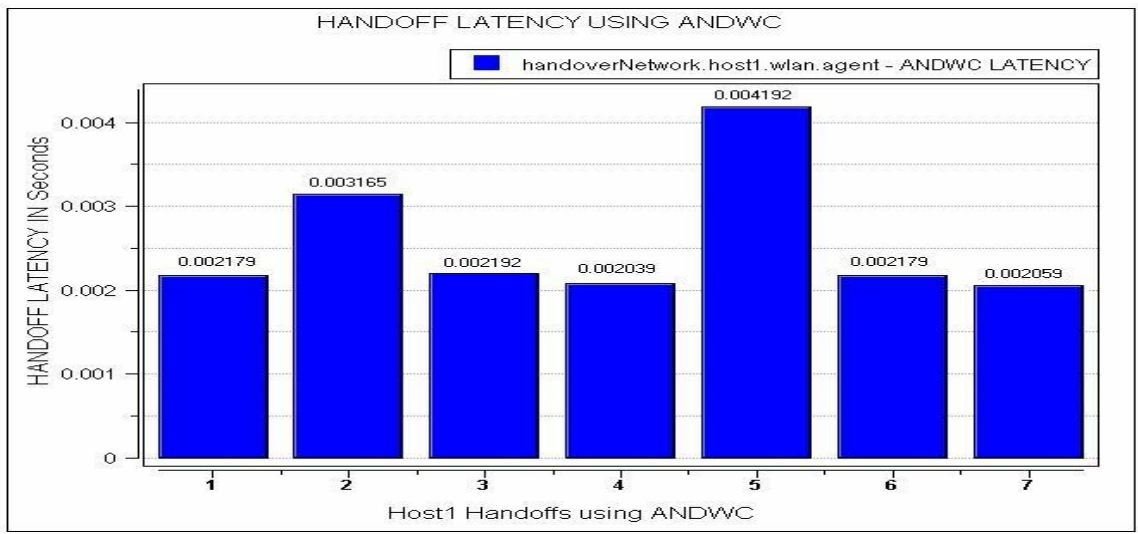

Figure 6. Handoff Latency Using Andwc